# Single-contact pressure solution creep on calcite monocrystals

Sergey Zubtsov[1,2], François Renard [*, 1, 3], Jean-Pierre Gratier[1],

Dag K. Dysthe[3] and Vladimir Traskine[2]

[1] LGIT-CNRS-Observatoire, Université Joseph Fourier, BP 53, 38041 Grenoble cedex,

France

Email: Francois.Renard@lgit.obs.ujf-grenoble.fr

[2] Colloid Chemistry Group, Chemistry Department, Moscow State University, Moscow,

119899 Russia

[3] Physics of Geological Processes, University of Oslo, Postboks 1048 Blindern, 0315 Oslo,

Norway

Corresponding author: François Renard

*Abbreviated title:* Pressure solution on calcite monocrystals

**Abstract:** Pressure solution creep rates and interface structures have been measured by two methods on calcite single crystals. In the first kind of experiments, calcite monocrystals were indented at 40°C for six weeks using ceramic indenters under stresses in the 50-200 MPa range in a saturated solution of calcite and in a calcite-saturated aqueous solution of $NH_4Cl$. The deformation (depth of the hole below the indenter) is measured *ex-situ* at the end of the experiment. In the second type of experiment, calcite monocrystals were indented by spherical glass indenters for 200 hours under stresses in the 0-100 MPa range at room temperature in a saturated aqueous solution of calcite. The displacement of the indenter was continuously recorded using a specially constructed differential dilatometer. The experiments conducted in a calcite-saturated aqueous solution of $NH_4Cl$ show an enhanced indentation rate owing to the fairly high solubility of calcite in this solution. In contrast, the experiments conducted in a calcite-saturated aqueous solution show moderate indentation rate and the dry control experiments did not show any measurable deformation. The rate of calcite indentation is found to be inversely proportional to the indenter diameter, thus indicating that the process is diffusion-controlled. The microcracks in the dissolution region under the indenter dramatically enhance the rate of calcite indentation by a significant reduction of the distance of solute transport in the trapped fluid phase. This result indicates that care should be taken in extrapolating the kinetic data of pressure solution creep from one mineral to another.

**Introduction**

Pressure solution creep (PSC), coupled with cataclastic and crystal plastic deformation, plays a significant part in sedimentary and fault rock deformation in the upper crust (Ramsay 1967). This is a mechanism of a slow aseismic compaction and porosity reduction whereby stress-enhanced chemical potential at the contact sites of mineral grains results in local supersaturation of the adjacent fluid, diffusion of the solutes out of the high concentration areas and precipitation of the material on the grain faces with smaller chemical potential (Weyl 1959; Paterson 1973; Rutter 1976). According to this definition, a distinction can be made between reaction-kinetics-controlled PSC, where the rate of deformation is limited by the dissolution or precipitation of the material, and diffusion-controlled PSC, where this rate is limited by solute diffusion along the grain contacts (Raj 1982).

The main difficulty to quantify which parameters control PSC rates is that this very slow mechanism of deformation is difficult to reproduce experimentally. Moreover, this process largely depends on the shape of the solid that evolves with time because of deformation. This is the main reason why indenter experiments have been developed; where an inert indenter in contact with a crystal and its solution is shown to pressure dissolve the crystal (Gratier 1993; Dysthe et al. 2003). As the shape of the indenter does not change with time, the deformation is easier to quantify. Moreover, the deformation laws for this simple geometry are well-known and allow separating the different parameters controlling the rate of deformation. By this means one expects to deliver creep laws for relevant geological conditions and to quantify both finite strain and rates of PSC deformation patterns.

The displacement rate of a piston ($\dot{\gamma}$) indenting a mineral by PSC is deduced from theoretical laws (Raj 1982; Rutter 1983) for these two types of PSC. For the diffusion-controlled case, the relationship is:

$$\dot{\gamma} \propto \frac{Dcw}{d^2} \tag{1}$$

where $c$ is the mineral solubility *at stressed state*, $w$ the thickness of the water film under the indenter, $D$ the diffusivity of rate limiting ion and $d$ the indenter diameter. Here,

$$c = c_0 e^{-\frac{\Delta\mu}{RT}} \qquad (2)$$

where $c_0$ is the mineral solubility at zero stress, $R$ the gas constant, $T$ the temperature and $\Delta\mu$ the difference in chemical potential between the dissolution and precipitation zones (proportional to the applied normal stress $\sigma_n$).

At low $\Delta\mu$ (if $\sigma_n <$ 30 MPa (Rutter 1976)) an approximation may be used to obtain the relation derived by Gratier (1993):

$$\dot{\gamma} \propto \frac{Dc_0 w \Delta\mu}{RTd^2} \qquad (3)$$

For the kinetic-controlled case, the relationship is:

$$\dot{\gamma} \propto k \frac{c - c_0}{c_0} \qquad (4)$$

or expressed more simply:

$$\dot{\gamma} \propto kc \qquad (5)$$

Here $\frac{c - c_0}{c_0}$ is the saturation state of the solution and $k$ the reaction rate constant of the interfacial reaction (dissolution or precipitation). For the precise calculations it should be borne in mind that this constant is temperature-dependent.

Using relation (2) at $\sigma_n <$ 30 MPa, the relation derived by Gratier (1993) for mineral indentation by kinetic-controlled PSC is obtained, namely:

$$\dot{\gamma} \propto \frac{kc_0 \Delta\mu}{RT} \qquad (6)$$

However, relation (4) is based on the approximation that the rate of interfacial reaction is proportional to the saturation state of the adjacent solution. This is not always true. In the general case, this rate (and thus the rate of mineral indentation by kinetic-controlled PSC) is in power relation with the saturation state (Morse & Arvidson 2002):

$$\dot{\gamma} \propto k(\frac{c - c_0}{c_0})^n \qquad (7)$$

At low applied stress, a relationship similar to that of de Meer & Spiers (1999) can be derived from (7):

$$\dot{\gamma} \propto \frac{kc_0 (\Delta\mu)^n}{RT} \qquad (8)$$

Here $n$ is the empirical reaction order that is different for different minerals. Moreover, it is different *for different solution compositions* and usually different for dissolution and precipitation processes (Morse & Arvidson 2002). For calcite in pure water, $n$ was found to be equal to 1 for dissolution (Plummer *et al.* 1978) and 2 for precipitation (Reddy & Wang 1980).

Generally, all of the preceding means that the relationship between the rate of mineral indentation by PSC and the value of the difference in chemical potential between the dissolution and precipitation zones (i.e. the value of applied normal stress $\sigma_n$) is sophisticated and depends on many factors. Care should be taken when using the chemical potential to determine whether kinetics or diffusion-controlled PSC takes place in an experiment. On the other hand, equations (1) and (5) show that, if the process of crystal indentation is diffusion-controlled, the rate of indenter displacement is inversely proportional to the square of the indenter diameter, and in the case of kinetic control the indenter diameter has no influence on this rate. The relationship between indenter diameter and indentation rate seems to be a more reliable parameter on which to base conclusions concerning the controlling mechanism of PSC.

Two types of laboratory experiment have been conducted previously to investigate PSC. In the single contact experiments, the crystal surface is indented by a piston or another crystal (Tada & Siever 1986; Gratier 1993; Dysthe *et al*. 2002). This type of experiment is conducted in order to observe the deformation process precisely controlling the applied normal stress, the contact surface area and the displacement of the indenter. These experiments represent a useful tool for validating existing theoretical pressure solution models (Dysthe *et al*. 2003). Unfortunately, for low-solubility minerals, such as calcite or quartz, the rate of single contact deformation is too slow and the laboratory experiment must last several months or the displacement recording system must be extremely sensitive. This is why a second experimental approach is used for these minerals. This consists of powder compaction, i.e. multiplying the number of grain contacts to increase the bulk deformation rate (De Meer & Spiers 1997; den Brok *et al*. 1999; Zang *et al*. 2002). However, the interpretation of the results obtained is ambiguous as other deformation processes (grain sliding, subcritical crack growth) can occur in addition to PSC (Zang *et al*. 2002).

Contrary to the significant amount of theoretical and experimental studies of cataclastic and plastic calcite deformation, the PSC of calcite has not previously been studied in great detail in spite of its importance in natural rock deformation processes. The main reason for this situation is the low solubility of calcite, which makes it difficult to investigate its PSC in laboratory conditions. However, a limited number of pressure solution compaction experiments using calcite powder have been performed. Baker *et al.* (1980) studied compaction of deep-sea carbonate Iceland spar and reagent-grade calcite powder at 35-100 MPa and 22-180°C. The duration of experiments varied from 21 to 240 hours. After 240 hours of compaction, a porosity reduction of 9.9% was obtained. Unfortunately, there is no clear evidence in these experiments that cataclastic and crystal plastic deformation do not make a significant contribution to the porosity reduction value obtained.

Recently, Zang *et al*. (2002) compacted fine-grained calcite powders at ambient temperature. The compaction experiments were performed in drained conditions with a saturated calcite solution, applying a uniaxial effective stress of 1-4 MPa. All samples were initially precompacted at 8 MPa to minimize the contribution of grain sliding and reorientation to subsequent wet deformation. The absence of brittle deformation during the experiment was proven by the absence of recorded acoustic emissions. However, a mechanism of subcritical crack growth was considered to be a relevant deformation process. Dry experiments and experiments with inert fluid did not show any significant deformation. Deformation of the wet samples reached 1.1% maximum strain after 20 days of compaction. The compaction rate decreased with increasing strain, increasing grain size and decreasing applied stress. On the basis of these data, it was concluded that reaction kinetic-controlled pressure solution creep and possibly subcritical crack growth are the dominant deformation mechanisms. To our knowledge, no single contact PSC experiments have been performed date because of the very slow rate of calcite single crystal deformation. There are at least three different approaches described below to resolve this problem: modifying the fluid chemistry, inducing microcracks, and increasing the sensitivity of the deformation measurement device.

*Effect of fluid chemistry*

The nature of the fluid is an important parameter for the rate of PSC. Above all, it is obvious that the rate of deformation depends on the solvent capacity of the fluid, i.e. it is proportional to the solubility of the mineral in the fluid. For example, Pharr & Ashby (1983) studied the rate of pressure solution deformation of pre-compressed potassium chloride and sucrose powders in methanol/water solutions. They found that the extent of deformation enhancement is a linear function of the solubility of the solid in the pore fluid. Gratier & Guiguet (1986) used a solution of NaOH to increase the deformation rate of quartz, starting from the idea that alkalinity enhances its solubility as well as its dissolution and growth kinetics. Moreover, the chemistry of the fluid can influence the rate of pressure solution deformation by phenomena such as adsorption of fluid components and complex formation. Skvortsova *et al.* (1994) investigated the effect of adding small quantities of dimethylaniline (DMA) on the rate of deformation of halite by PSC. They found that DMA forms complexes with salt and adsorbs on the salt crystal surfaces, resulting in a change from a diffusion-controlled to a reaction-controlled mechanism. This transition is accompanied by a sharp decrease in deformation rate. In addition, Zang *et al.* (2002) found that calcite powder PSC slows down significantly if a small quantity of $MgCl_2.6H_2O$ is added to the pore fluid. The reason for this effect is the adsorption of magnesium ions on the calcite crystal surfaces, which inhibits further calcite crystallization.

The rate of calcite deformation by PSC can also be enhanced by changing the fluid chemistry. As calcite is a weak base, its solubility is increased in acid solutions. In particular, in weak acid solutions, a dynamic chemical equilibrium takes place so that calcite can not only be dissolved in the solutions, but *also precipitate* if the solution is supersaturated. This is not the case for a total dissolution reaction, as for example when calcite dissolves in HCl. A well-known natural example of this dynamic equilibrium is the dissolution of calcite in a solution saturated in carbon dioxide:

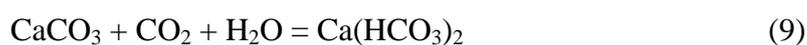
$$CaCO_3 + CO_2 + H_2O = Ca(HCO_3)_2 \qquad (9)$$

The rate of mineral deformation by PSC is proportional to its solubility; consequently the rate of calcite deformation by PSC should be enhanced in a weak acid solution saturated by calcite.

*Effect of microcracks*

Gratier *et al.* (1999) have shown in experiments of halite indentation by diffusion controlled PSC, that if the applied stress is high enough to form microcracks on the mineral surface under the indenter, the rate of deformation enhances dramatically. The reason for this effect is that the kinetics of diffusion along a trapped fluid phase under the indenter could be one order of magnitude less than in bulk fluid (Dysthe *et al.* 2002; Alcantar *et al.* 2003). If there are no microcracks under the indenter, the indentation rate is inversely proportional to the square of the indenter diameter. However, in the presence of microcracks, faster diffusion out of the contact may occur in the free fluid that fills the microcracks. Consequently, in this case the rate of mineral indentation is not limited by the diameter of the indenter, but by the distance between two adjacent microcracks. It should be noted that calcite is a brittle mineral and the formation of microcracks on its surface under the indenter seems plausible.

*Increase in recording sensitivity*

Another possibility for carrying out a single-contact PSC experiment with calcite is to increase the sensitivity of the displacement recording system. Dysthe *et al.* (2002, 2003) developed a capacitance dilatometer especially designed for single-contact pressure solution experiments. The sensitivity of this device is about $10^{-9}$ m. Halite monocrystals were tested using the dilatometer and showed a creep rate of about $2 \times 10^{-6} – 10^{-8}$ m/day under an applied stress of 4 MPa (Dysthe *et al.* 2003).

In this contribution, two types of single-contact PSC experiments with calcite monocrystals are described. The three above-mentioned ways of activating PSC on calcite crystals were successfully used in this study. The interpretation of the results obtained provides new constraints for understanding the mechanism of PSC on calcite deformation.

**Experimental methods**

*PSC indenter experiments with ex-situ deformation measurement*

With the simple indenter technique described in detail by Gratier (1993), calcite monocrystals were indented at 40°C using ceramic indenters under stresses in the 50-200 MPa range for six weeks in a saturated solution of calcite and in a calcite-saturated aqueous solution of $NH_4Cl$ (Fig. 1a).

Natural calcite crystals (from Sweetwater Mine, Reynolds Country, Missouri, USA, delivered by Ward's Natural Science Establishment Inc., Rochester, NY, USA) were cleaved and then polished (1200 to 4000 mesh grinding paper and DP-mol cloth with 1 micron diamond suspension, all from Struers A/S, Rodovre, Denmark). The surfaces were cleaned with water. The aqueous solutions were checked to ensure that they wetted the crystals when the experiments were started. For some experiments without water, the calcite crystals were dried with nitrogen. For each experiment a different crystal was used, and the indenter was always applied on the same crystallographic face.

The saturated solution of calcite was prepared by adding several grams of $CaCO_3$ powder to 100 ml of distilled water boiled for 15 minutes before solution preparation in order to eliminate any traces of $CO_2$, and heated to 60°C, shaking once per day for several days. The aqueous calcite saturated solution of $NH_4Cl$ was prepared by adding 10 g of $CaCO_3$ to 100 ml of a 5% solution of $NH_4Cl$ in distilled water and heating to 60°C, shaking once per day for several days. Before starting the experiments, the solutions were kept at 40°C (temperature of the experiments) for two days.

Each deformation cell consists of a Plexiglas tube with a diameter of several centimetres, and a Plexiglas bottom. The ceramic indenters, 300 µm wide and of various lengths (see Table 1), are mounted under a free-moving Teflon piston and put in contact with calcite monocrystals immersed in a saturated solution. Dead-weights placed on the piston maintain a constant stress on the samples. The value of the dead-weight is calculated for each indenter in order to impose the desired stress (Table 1). Liquid paraffin is added on the top of each cell to prevent evaporation of the fluid. The cells are placed in a furnace at constant temperature. After six weeks, the depths of the resulting holes are measured. Experimental conditions (stress, length of indenter, nature of fluid) for different deformation cells, together with the depth of the holes obtained are presented in Table 1. The indenter lengths are different for each indenter (from 300 to 740 μm), but the diffusion distance of solutes under the indenter is always the same because of the same indenter widths (300 μm).

*High resolution PSC experiments with continuous deformation recording*

Experiments were conducted using the differential dilatometer described in Dysthe *et al*. (2002, 2003) which allows displacement to be measured accurately with a long-term resolution of 2 nm (Fig. 1b). The device consists of two nominally identical capacitances, indenters and samples, one wet and one dry. This symmetrical design serves to eliminate all effects except those caused by differences on the two sides (presence/absence of water). The temperature is controlled by thermostatically-regulated water flowing through brass tubes spiraling down the external part of the instrument. The electronic circuit is a specialized lock-in amplifier built in-house and based on that of Jones (1988).

About 20-30 glass spheres with diameters ranging from 250 to 300 μm were glued along the edge of a cylindrical glass support. This kind of indenter geometry provides good mechanical stability during the experiment. As the experiment goes on, the spheres penetrate into the 1.5 mm diameter calcite crystal. The contact surface area between the indenter and the crystal gradually increases, resulting in a progressive decrease in stress.

The samples and solutions were prepared as described for the long-term calcite monocrystal indentation experiments.

Table 2 shows the experimental conditions for these experiments. All experiments were run at stabilized room temperature. The temperature was controlled with circulating water from a thermally-controlled bath, with fluctuations of about +/-0.1°C, that are larger than that described in Dysthe *et al*. (2003) for pressure solution on halite. Because the solubility of calcite is lower than halite and not very dependent on temperature (0.07% change with 0.1°C temperature change) the calcite experiments are not as sensitive to temperature fluctuations as the halite experiments.

From time to time the electrical noise in the environment of the instrument causes short fluctuations in the measured indentation. These fluctuations make quantitative interpretation difficult for some parts of the experimental curves.

**Results**

*PSC indenter experiments with ex-situ deformation measurement*

Fig. 2 shows a typical hole obtained after calcite monocrystal indentation. The depth of holes was measured after the experiments using a microscope with a resolution of 3 microns. The results are presented in Table 1. This table shows that none of the dry experiments (experiments Nos. 8, 9, 28, and 29) show any trace of deformation even under a maximum stress of 200 MPa. The experiments conducted with saturated calcite solution also usually do not show any deformation or show imprints of no more than 5 µm depth after six weeks (experiments Nos. 1, 3, 4, 6, 7, 11-13, 15-20). Depths greater than 5 µm have been observed in two experiments only (10 and 16), which can be interpreted as brittle deformation. Therefore, it may be concluded that, under these conditions, there is no significant calcite deformation if saturated calcite solution is used as a fluid.

On the other hand, the experiments conducted in a calcite-saturated aqueous solution of $NH_4Cl$ show significant deformation (Table 1, experiments Nos. 21-24, 26, 27, 30-35, 37-38, 40). Moreover, these data indicate a correlation between the depth of the hole and the applied stress (Fig. 3).

*High resolution PSC experiments with continuous displacement recording*

    *Experimental data*

Fig. 4 shows the results of the first PSC experiment with the double capacitance dilatometer. During this experiment the load was varied as shown in Fig. 4a. In this experiment the vertical displacement of the indenter is recorded continuously. This displacement corresponds to the indentation of some glass spheres in the calcite crystal. The temperature during the experiment is shown on Fig. 4b.

Figs. 4a, b show that differential measurement makes the double capacitance insensitive to temperature fluctuations of the order of 0.1°C. The disturbances in the indentation curve during the first five hours of the experiment arise from the failure of the temperature control system (see Fig. 4b).

Fig. 4a shows that there are large short-term fluctuations in the measured indentation. These are caused by electrical noise in the environment of the instrument. However, it is possible to separate the rapid jumps due to noise from the slowly varying capacitance due to indentation by PSC.

Table 3 shows the indentation rates obtained by linear regression to the noise-free periods of the indentation curve shown in Fig. 4a. The rate varies from 6 to 47 nm/h. It is believed that these variations represent the changes in stress under the indenter and the dissolution of minor asperities of calcite. However, the most important result of this experiment is the possibility of measuring the rate of calcite deformation with a resolution never reached before and to obtain relevant indentation rates.

As illustrated in Fig. 4c, an increase in dead-weight by a factor of 2.1 (from 80 to 167 grams) has an immediate effect on the deformation rate which increases by a factor of 2.7 (from 11 nm/h in the interval 38-46 h to 30 nm/h in the interval 46-54 h). Fig. 4d also shows that if the dead-weight is removed, the indentation process stops completely. The electrical disturbances make it difficult to view this effect clearly throughout the whole stress-free period. Nevertheless, the moderately long interval without electrical disturbances and without any movement of the indenter can be distinguished in Fig. 4d (Table 3, interval 109.5-113.3 h). The indentation process starts again with the same rate as soon as the dead-weight is replaced in its original position (Table 3, intervals 79-89 and 122-128 h).

The results of the second experiment are shown in Fig. 5a. The only difference from the first experiment is a new indenter with a different number of spheres touching the crystal surface. Fig. 5b shows the temperature of the experiment.

The rates of indentation in the noise-free parts of the signal are 48 and 56 nm/h at the beginning and end of the experiment, respectively.

Fig. 6a shows the results of calcite dry indentation. The indentation curve only shows some electrically-induced fluctuations between 0 and 200 nm, which are 10-15 times less than the deformation values obtained in the wet experiments for the same period of time.

*Profiles of indented regions*

The sample surfaces with indented regions are imaged by white light interferometer (WLI) microscopy (Wyko 2000 Surface Profiler from Veeco). The profiler is a microscope with a reference arm creating interference fringes with maximum intensity at equal optical path lengths of the imaging beam and reference beam. By vertical movement of the sample and simultaneous image capturing, the interference intensity envelope and thus the relative height of the imaged surface at each pixel is determined with a resolution of 3 nm. The horizontal resolution depends on the lens used and with the highest magnification it is at the diffraction limit of about 0.5 μm.

The calcite crystal used in the first experiment was examined using this method. Fig. 7a shows an indented part of a crystal surface. This image shows all the spherical imprints of the indenter and the scratches produced by the horizontal movement of the indenter. The depth of the holes is about 0.65 μm and that of the scratches is less than 0.30 μm. These observations show that the change in dead-weight during the experiment causes minor displacement of the indenter accompanied by surface scratching. When the spheres are placed under stress again, they restart crystal indentation until a new change in dead-weight. This is the reason for the moderate depth of the holes measured by WLI, comparable to the values of deformation obtained in the experiment.

Fig. 7b presents a typical spherical dissolution imprint and its 2D profile. The depth of the hole is about 1 μm and its diameter is approximately 50 μm. The hole is surrounded by rings, which are about 200 nm higher than crystal surface. These rings represent the reprecipitated calcite. Observations also indicate that the holes are produced by the dissolution of calcite and not by subcritical crack growth and consequent brittle deformation. There is also evidence that some cracks formed under the indenter along the calcite cleavage planes. These cracks are produced by the action of high stress on the brittle mineral.

*Stress estimation*

For further calculations, the stress applied on the crystal surface in our experiments, should be estimated. Given that the dead-weight is 167 g, the diameters of the spherical imprints on the crystal surface is about 50 μm, the number of spheres touching the crystal surface is around 10, and the stress could be assessed at 100 MPa.

**Discussion**

*Dominant deformation mechanism*

*PSC indenter experiments with ex-situ deformation measurement*

The absence of any imprints on the crystal surface in the dry experiments strongly suggests that there is no crystal plastic deformation of the crystals and that the dominant mechanism is water-assisted deformation. The presence of dissolution imprints suggests that PSC is the main deformation mechanism.

*High resolution PSC experiments with continuous deformation recording*

The deformation process stops in the absence of dead-weight, i.e. when no more stress is applied to the crystal surface. This fact indicates that, in the experiments conducted, deformation is effectively being measured and not electrical fluctuations. The spherical imprints of the indenters on the crystal surface (Fig. 7) indicate that the deformation measured is the deformation of the crystal surface itself and not of parts of the experimental device. The absence of any significant deformation in the dry experiments strongly indicates that the dominant mechanism of deformation is pressure solution. The presence of nearby grown calcite around the indentation imprint (Fig. 7b) also confirms this conclusion.

*Rate limiting process*

The rate of calcite monocrystal indentation in the high resolution PSC experiments is about 1 μm/day for a stress of 100 MPa (see Table 2 and the results of the second experiment). On the other hand, the experiments with *ex-situ* deformation measurements did not show any significant deformation after 42 days of indentation, even under a maximum stress of 200 MPa. The difference between these two types of experiment is the length of the diffusion path under the indenter which is 6 times shorter in the high resolution PSC experiments than in the experiments with *ex-situ* deformation measurement (the diameter of the spherical imprint on the crystal of the high resolution experiment is 50 μm compared to the 300 μm width of the indenters in experiments with *ex-situ* deformation measurement). Given that the rate of crystal indentation is independent of indenter diameter in the case of kinetic-controlled PSC (equation 4) and inversely proportional to the square of the indenter diameter in the case of diffusion-controlled PSC (equation 3), it could be suggested that diffusion is the limiting step in our experiments. To test this assumption, the rate of crystal indentation in the experiments with *ex-situ* deformation measurement should be calculated for the diffusion-controlled case on the basis of the rate in the high resolution experiments. Starting from the difference in the length of diffusion paths, it is easy to see that this rate should be about 0.03 μm/day multiplied by a factor that represents the difference in the stress in these two experiments. This means that the final deformation after the end of the experiment (42 days) should be no more than several microns. This value is in good agreement with the experimental result – the absence of any measurable deformation after the end of the experiment. Thus it may be concluded that calcite indentation by PSC under our experimental conditions is limited by the diffusion of solutes under the indenter.

The diffusion control of calcite PSC in our experiments is a crucial difference between our results and those of Zang *et al.* (2002). The plausible explanation of this difference is the indenter technique used here compared to the powder compaction method. Because of the very slow rate of calcite deformation by PSC, the maximum value of powder compaction obtained was only 1.2%. This means that the contact interface areas between two adjacent grains stay very small and irregular throughout the experiment. Therefore, the diffusion distances in this experiment are always extremely short. This may be the reason for kinetic control operating in the calcite powder compaction experiments. On the other hand, in the indentation experiments, significantly larger diameters of indenter impose a longer diffusion path, and the process is thus diffusion controlled.

*Effect of the mineral solubility*

The experiments with *ex-situ* deformation measurement with a saturated aqueous solution of calcite do not show any significant deformation. On the other hand, in the experiments conducted in a calcite-saturated aqueous solution of $NH_4Cl$, deformation of several tens of microns can be observed (Table 1). The reason for this change probably is the effect of ammonium chloride as a weak acid on the solubility of calcite. Solutions of weak acids should increase the rate of calcite deformation by PSC. For example, in an aqueous solution of $NH_4Cl$ saturated by calcite there is a dynamic equilibrium:

$$2CaCO_3 + 2NH_4Cl + 2H_2O = Ca(HCO_3)_2 + CaCl_2 + 2NH_4OH \quad (10)$$

The chemical potential of the $CaCO_3$ surface under the indenter is higher than that of the free surface. Therefore, in this region, the equilibrium is displaced to the right hand side of equation (10) and dissolution of calcite takes place. This additional quantity of dissolved calcite diffuses to the free surface, where the concentration of dissolved calcite is lower, and precipitates there. The solubility of calcite in a 5% solution of $NH_4Cl$ is about 0.5 g/l (Lafaucheux 1974), that is 30 times higher than in pure water. This means that in this solution the final depth of the holes obtained after the end of experiment should be 30 times higher than in the case of pure calcite solution. Based on the depth value of several microns in the case of pure calcite solution (see calculations in previous section), the depth of holes in the presence of $NH_4Cl$ may be estimated to be several tens of microns. This value is in agreement with the experimental results.

*Effect of stress*

*PSC indenter experiments with ex-situ deformation measurement*

Unfortunately, the scatter of the data (Fig. 3) makes it impossible to obtain a good approximation of this correlation by any simple function. On this figure, a straight line is represented as a guide for the eyes, but a power law or an exponential function could also fit the data. It may be only noted that enhancement of the applied stress increases the rate of mineral indentation by PSC.

*High-resolution PSC experiments with continuous deformation recording*

Fig. 7a shows that the change in dead-weight during the experiment causes some minor horizontal displacement of the indenter. When the spheres are under stress again, they restart crystal indentation until a new change in dead-weight. During the first minutes of indentation, the contact area between the spheres and the surface is extremely small because of the small irregularities that are present even on a very well polished surface; therefore the spheres apply an excessively high stress on new areas of the crystal surface. As a consequence, a dramatic increase in deformation rate should be detected during the first minutes after the change in dead-weight. This effect is clearly seen on the experimental curve when the dead-weight changes from 0 to 167 g at t=144h (Fig. 4d). On the other hand, this is not the case when the dead-weight changes from 80 to 167 g at t=46h (Fig 4c). Maybe, in this second example, the indenter did not move horizontally during the change in dead-weight and that each sphere stayed in its initial hole.

An increase in dead-weight by a factor of 2.1 has an immediate effect on the rate of deformation which increases by a factor of 2.7 (Fig. 4c). We propose that in our experiment, the rate of mineral deformation is almost directly proportional to the applied stress. Note that we only have one data to support this conclusion.

*Comparison with PSC on halite*

*PSC indenter experiments with ex-situ deformation measurement*

The experiments with a saturated solution of calcite in 5% solution of $NH_4Cl$ show indentation rates of about 1 µm/day (Table 1). This value is the same as in the experiments of Gratier (1993), where the rate of halite monocrystal deformation in the same type of experiments with diffusion control was found to be about 1 µm/day at the applied stress of about 50 MPa and with an indenter diameter of 200 µm. Given that the solubility of calcite in a $NH_4Cl$ solution is 700 times lower than that of halite, that the rate of calcite indentation should also be slowed down by a factor of 2 compared to halite because of the difference in the indenter diameters (300 µm in calcite experiments and 200 µm in halite ones), and that this rate should be enhanced several times because of the higher applied stress used in our experiments (the rate of mineral deformation by diffusion-controlled pressure solution is proportional to the applied stress in the case of moderate stress and in exponential proportion to the stress if the stress is high), it can be estimated that calcite deforms two orders of magnitude faster than it should when considering the difference in solubility between calcite and halite.

*High resolution PSC experiments with continuous deformation recording*

The observed rate of calcite indentation is of the order of several tens of nm per hour (see Table 3). It should be noted that this rate does not change significantly during the experiment. This is an important difference between our results and those described in Dysthe *et al*. (2003) for similar experiments with halite single crystals with diffusion control, where the indentation rate varies by several orders of magnitude throughout the experiment: from 2 µm/h at the start, when stress is imposed, to 2 nm/h after several days. However, on most of the calcite indentation curve (from 10 to 100 h) the indentation rate falls in the range between 27 and 180 nm/h (Table 2). These data show that calcite deforms only 5 to 10 times slower than halite, whereas its solubility is 20 000 times lower.

For our experiments, the stress may be estimated at 100 MPa. This stress is 25 times higher than that used in the halite experiments. This difference enhances the rate of mineral indentation by two orders of magnitude. In addition, the difference in the lengths of diffusion paths (50 μm in our experiments compared to 30 μm in the halite indentation experiments) slows down this rate by a factor of 2.5. However, even with these explanations, the rate of calcite deformation remains two orders of magnitude higher than it should be if compared with halite when considering the difference in solubility.

*Comparison between the two types of experiments*

If this method is used to compare the results of Gratier (1993) and Dysthe *et al.* (2002, 2003) for the rate of halite indentation, it would be found that these two rates are of the same order of magnitude provided that they are scaled with respect to the stress and the indenter diameter. On the other hand, for the two types of calcite experiments, the deformation rate is two orders of magnitude higher than it should be when compared with halite considering the difference in solubility. Consequently, there are some additional effects that may be responsible for the fast rates measured on calcite.

*Effect of micro-cracks*

For an additional explanation, the structure of the crystal surface should be considered. It is known that in diffusion-controlled PSC, the presence of microcracks in the region of dissolution under the indenter dramatically enhances the rate of mineral indentation by a significant reduction in solute transport distances in the trapped fluid phase (Gratz 1991; den Brok 1998; Gratier *et al.* 1999). The reason for this effect might be related to the substantial difference in transport properties between the trapped fluid phase along the grain contact and the bulk fluid (Rutter 1976). Because of this, the diffusion flux in microcracks (bulk fluid) is greater than along the trapped fluid phase. If the dissolution surface is cut by microcracks, the average diffusion distance in the trapped fluid phase is significantly reduced and the deformation rate is increased dramatically (den Brok 1998). This effect is believed to be critically important in our experiments, as a considerably high stress is applied on a brittle mineral. Cracks seen on the interferometry images of the indenter imprints in the high-resolution PSC experiments provide support for this assumption. The wide scatter of hole depths obtained in the same experimental conditions (Fig. 3) also confirms the microcrack hypothesis. The microcrack structure is unique for each experiment, which is the reason for the significant variation of indentation rates. Precise understanding of this interface structure needs further investigation.

Microcracks might be formed by the release of elastic forces when the indenter is left on the crystal. Another effect is the formation of twinning that can be readily achieved in laboratory conditions. Twinning could lead to localized hardening and could promote further fracturation. This is a fast process that potentially happens in the early stages of the deformation process and controls the initial fracture spacing.

At this stage, it can only be unequivocally assessed that the calcite interface structure under the indenter is radically different compared to that of halite. This result indicates that great care must be taken in extrapolating the kinetic data of pressure solution creep obtained from one mineral to another.

**Conclusion**

This article reports on the first experimental study of calcite monocrystal indentations by the pressure-solution creep.

- The rate of mineral deformation was found to be inversely proportional to the square of the indenter diameter, which means that the process is diffusion-controlled.
- When comparing the experiments with various calcite solubility (aqueous solution versus solution of $NH_4Cl$), the rate of calcite indentation was found to be proportional to its solubility.

- The rate of calcite deformation was found to be almost proportional to the applied stress.
- The observed rate of calcite indentation is two orders of magnitude higher than it should be when compared with the halite indentation rate in the same type of experiments, taking into account the difference in solubility of these two minerals. This effect could be explained by the presence of microcracks under the indenter phase. Understanding of this effect needs further studies of the interface structure.

Our experimental data suggest that PSC deformation on calcite can be quite fast. This process should occur in nature (basin compaction, fault gouge) with a wide range of rates, depending on stress, fluid composition and crack density. The presence of microcracks should enhance PSC in active faults after an earthquake as new surfaces are created and new "indenters" are formed. This could explain the fast fault creep observed after a major earthquake (Donnellan & Lyzenga 1997).

The project has been supported by the ANDRA, the CNRS (Action Thématique Innovante), the Norwegian Research Council through the *Fluid Rock Interaction* Strategic University Program (grant 113354-420), and the Center for Advances at the Academy of Science of Norway. We would like to thank R. Guiguet, D. Tisserand, C. Pequegnat and L. Jenatton for their technical help, and D. Gapais, S. Reddy and B. den Brok for their constructive reviews.

**Captions of figures**

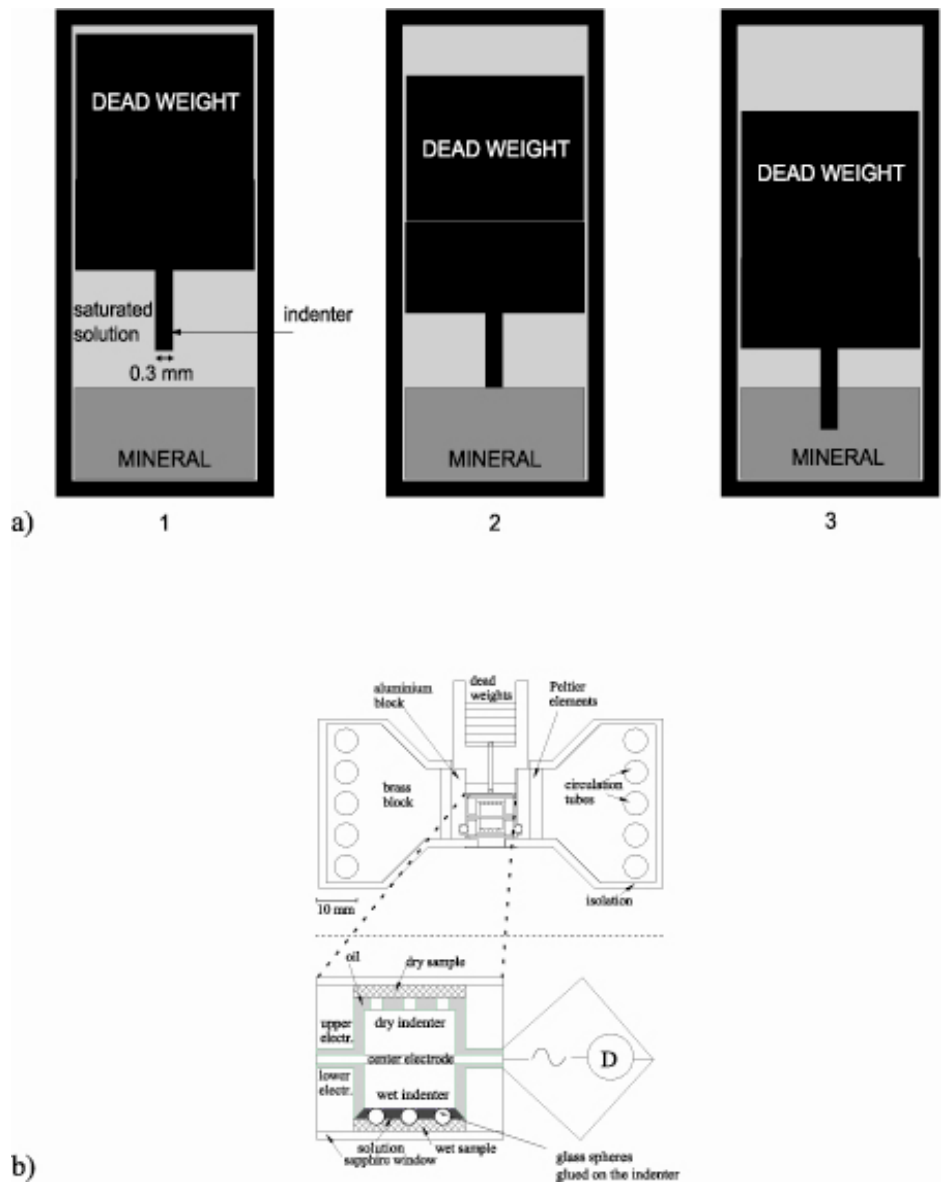

Fig. 1. (a) PSC indenter experiments with *ex-situ* deformation measurement. First stage: equilibrium between calcite crystal and its saturated solution. Second stage: the indenter is placed on the mineral surface and vertical stress is applied through a dead-weight. Third stage: slow creep of the indenter downwards into the crystal. The deformation is analyzed *ex-situ* by measuring the depth of the imprint. (b) High-resolution PSC experiments with continuous displacement recording through a double capacitance device (Dysthe *et al.* 2002, 2003). The technique allows displacement to be measured accurately with a long-term resolution of 2 nm (see text).

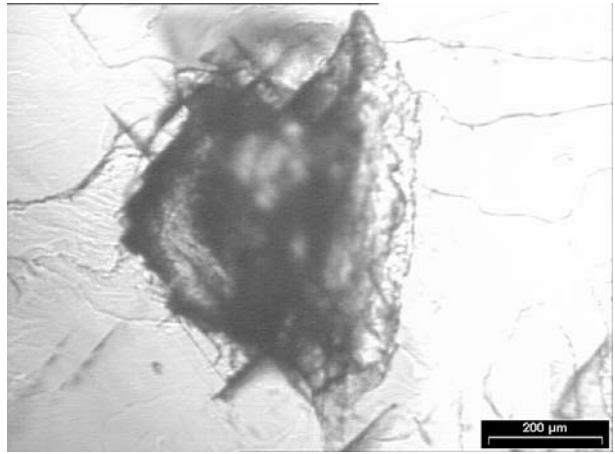

Fig. 2. Typical hole in a calcite monocrystal obtained after six weeks of indentation. The shape of the hole, close to a parallelepiped, represents that of the indenter. The depth of each hole is measured under microscope at the end of the experiment.

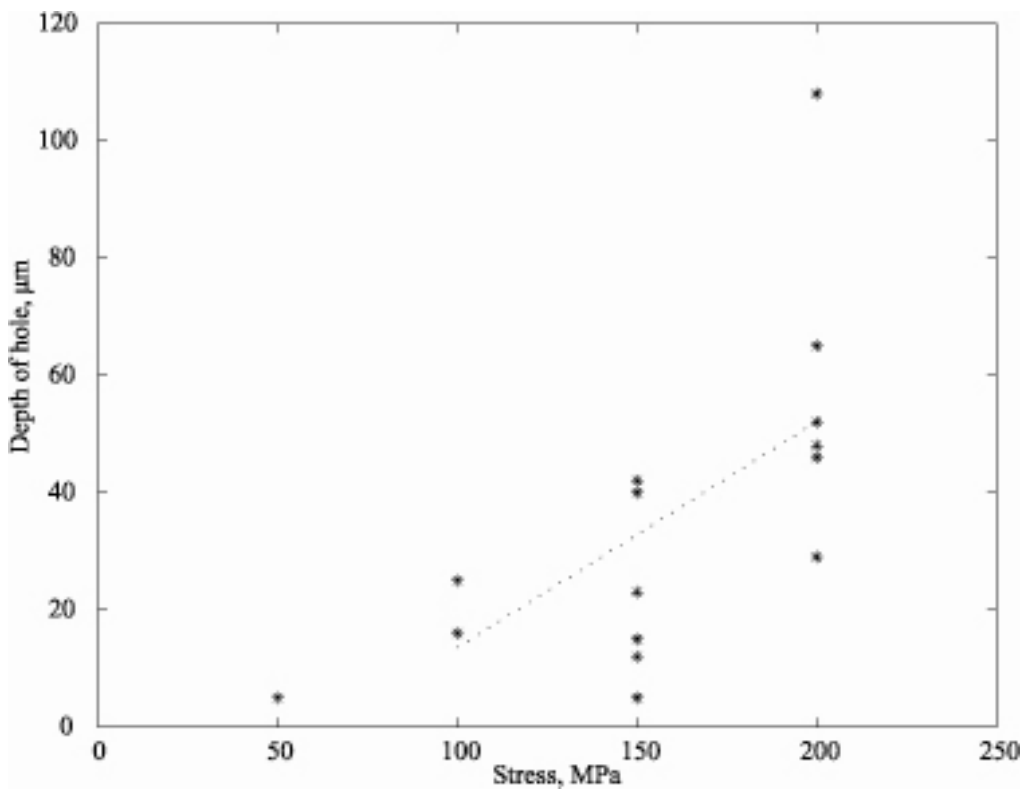

Fig. 3. Relationships between depth of hole and applied stress in the long-term indentation experiments conducted in a calcite-saturated aqueous solution of $NH_4Cl$. The straight line represents a guide for the eye.

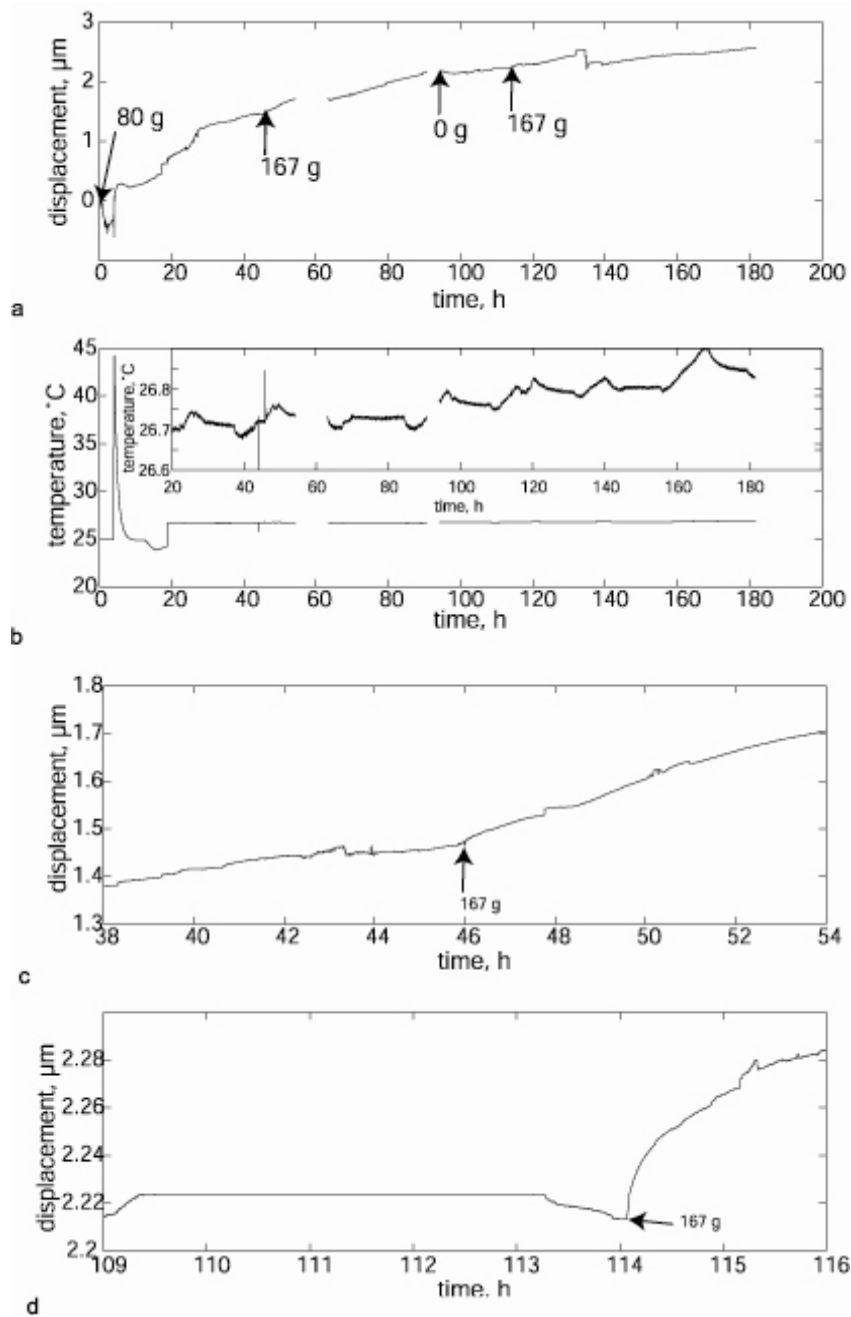

Fig. 4. Experimental results of the first high-resolution pressure solution creep experiment. (a) Indentation curve. The dead-weight was changed several times in the course of the experiment as shown on the figure. (b) Temperature during the experiment. (c, d) Zooms on the creep curve.

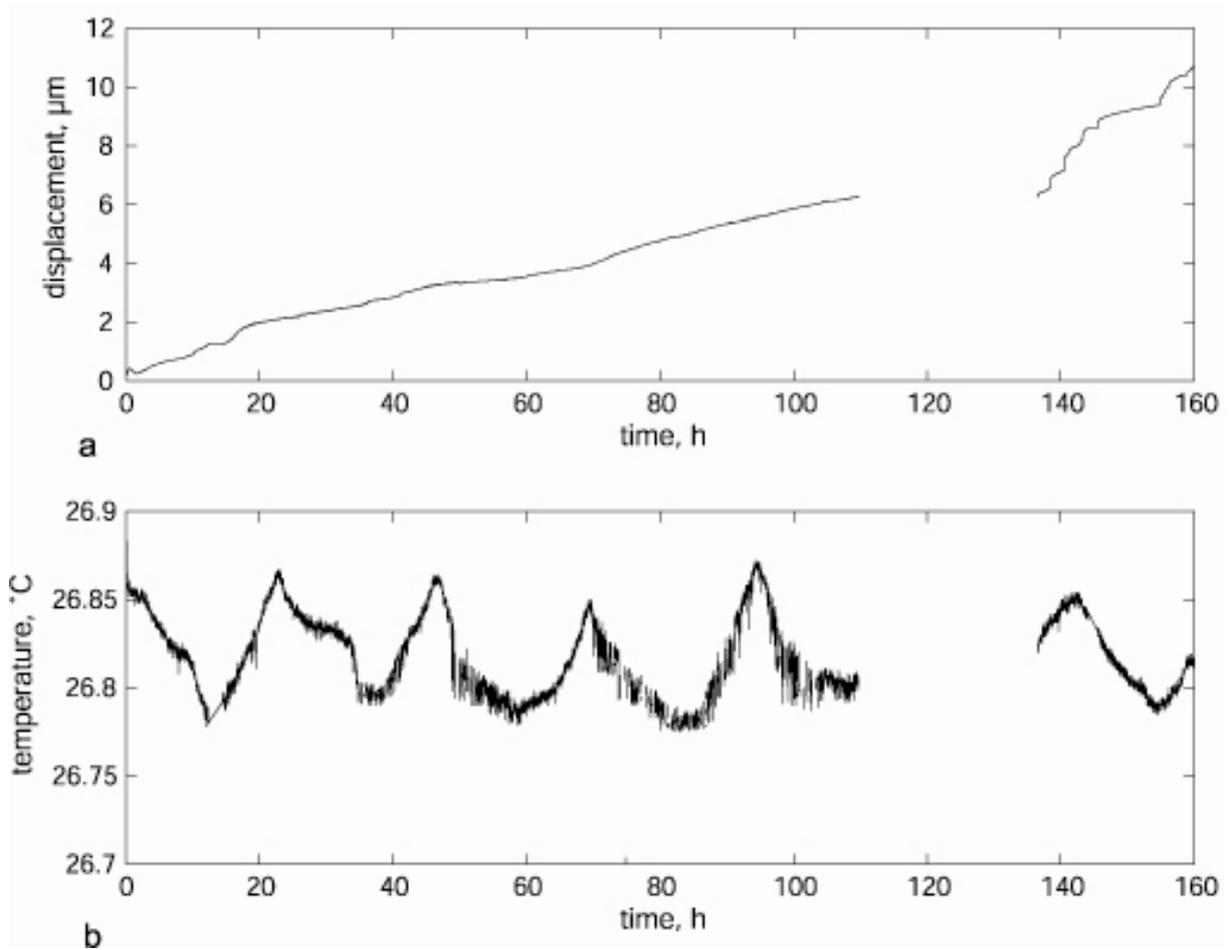

Fig. 5. Experimental results of the second high-resolution pressure solution experiment. (a) Indentation curve. (b) Temperature during the experiment.

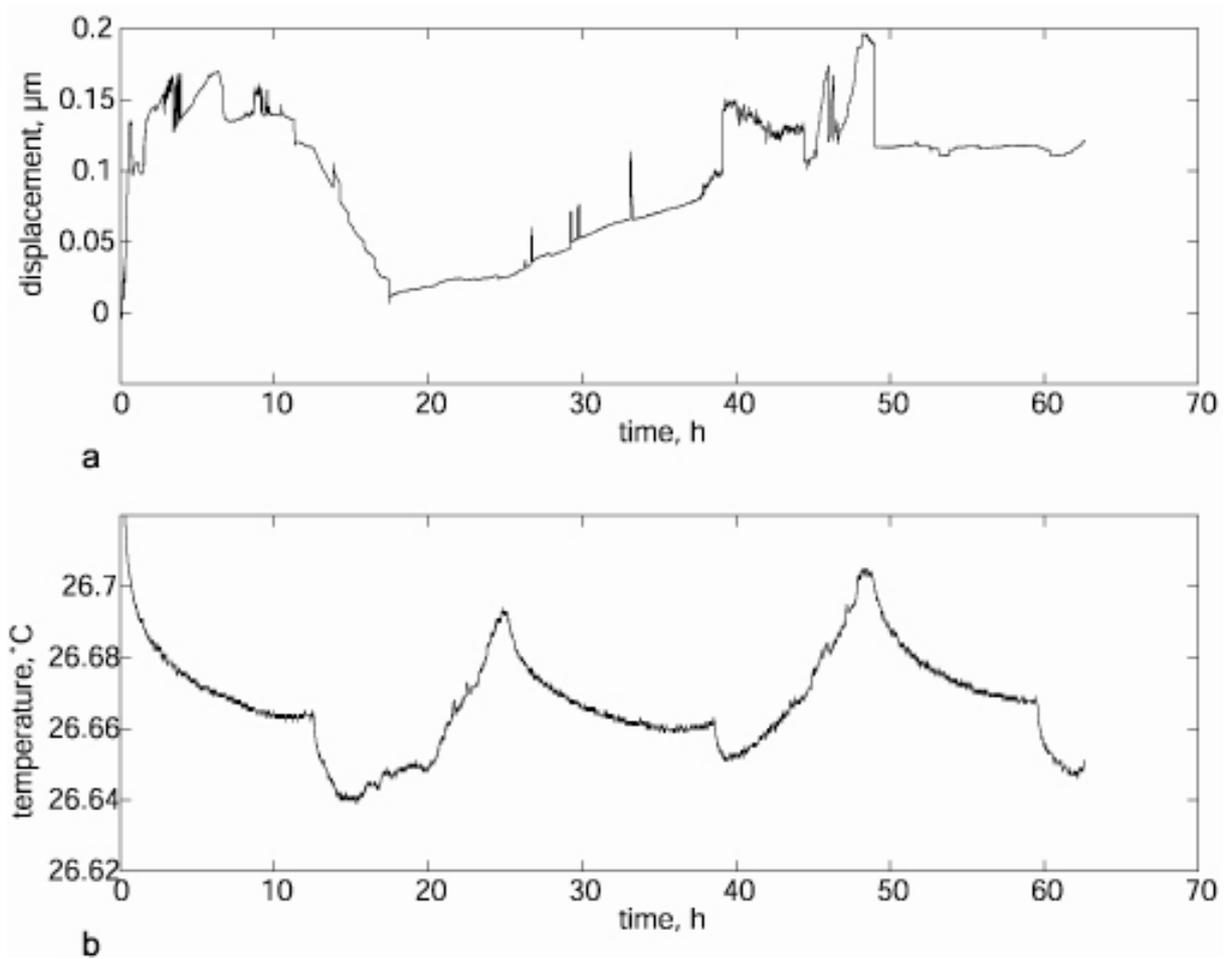

Fig. 6. Experimental results of the third (dry) high-resolution pressure solution experiment. (a) Indentation curve. (b) Temperature during the experiment.

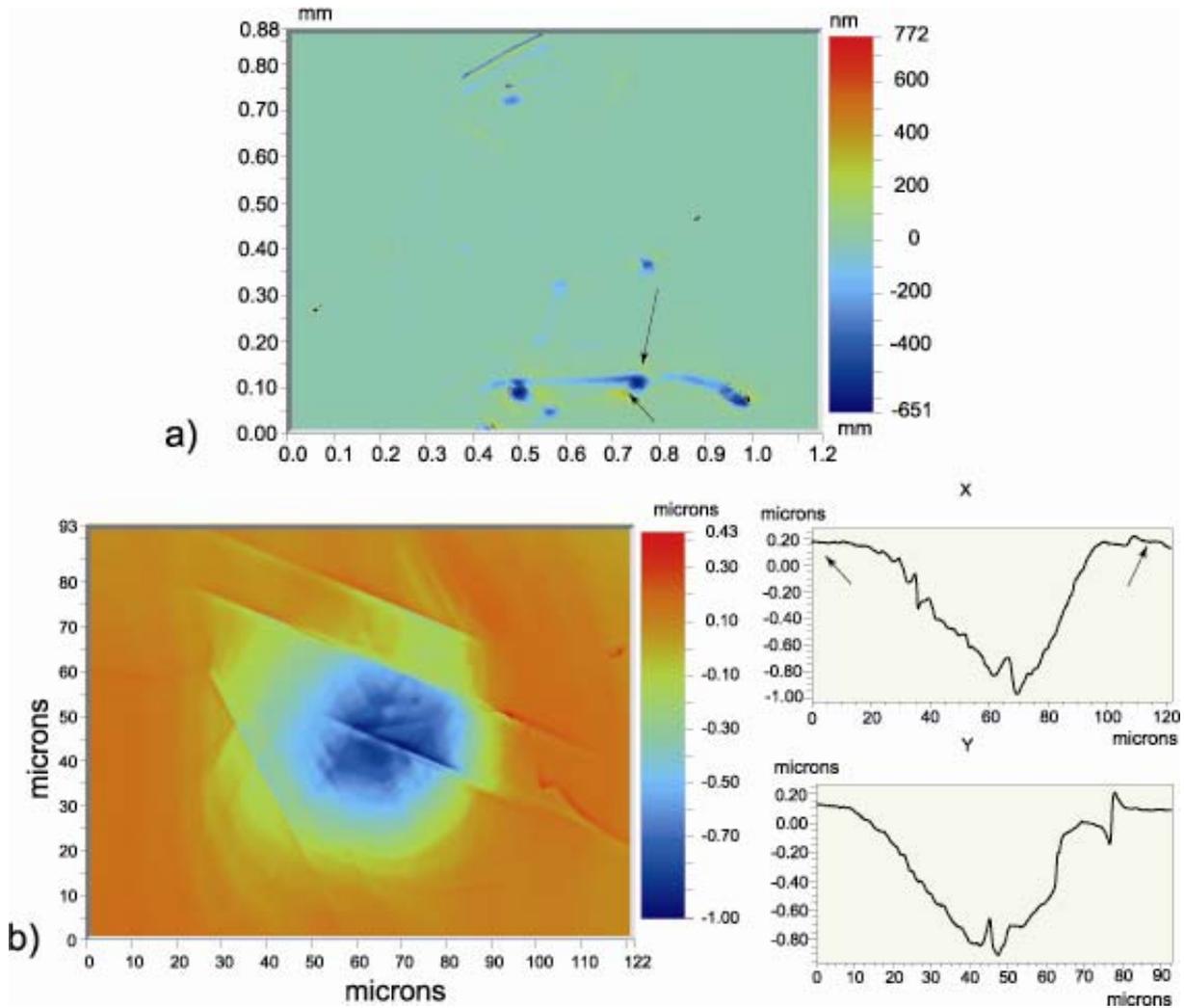

Fig. 7. White light interferometry image of the indented surface of a calcite monocrystal from the first high-resolution pressure solution creep experiment. (a) Spherical imprints of the indenter and some scratches produced by the horizontal movement of the indenter on the whole calcite crystal. (b) Zoom on a typical imprint of a sphere on the calcite surface and its cross-sections, which can be used to determine the depth of the imprint and also shows the 200 nm height rims of calcite precipitation near the edge of the indentation (arrows).

Table 1. Experimental conditions and results for PSC indenter experiments with ex-situ deformation measurement. All the experiments last for six weeks at constant temperature (40°C). All the indenters have a constant width (300 μm) and varying lengths

| Nb. | Length of the indenter (μm) | Dead-weight (kg) | Stress (MPa) | Solution chemistry | Depth of hole (μm) |
| --- | --- | --- | --- | --- | --- |
| 1 | 300 | 1.20 | 200 | calcite | 0 |
| 3 | 380 | 1.52 | 200 | calcite | 0 |
| 4 | 360 | 1.44 | 200 | calcite | 0 |
| 6 | 310 | 1.24 | 200 | calcite | 5 |
| 7 | 320 | 1.28 | 200 | calcite | 5 |
| 8 | 500 | 2.00 | 200 | dry | 0 |
| 9 | 340 | 0.77 | 150 | dry | 0 |
| 10 | 300 | 0.68 | 150 | calcite | 19 |
| 11 | 400 | 0.90 | 150 | calcite | 0 |
| 12 | 400 | 0.90 | 150 | calcite | 5 |
| 13 | 300 | 0.68 | 150 | calcite | 5 |
| 14 | 320 | 0.72 | 150 | calcite | 18 |
| 15 | 300 | 0.68 | 150 | calcite | 5 |
| 16 | 300 | 0.68 | 150 | calcite | 0 |
| 17 | 300 | 0.30 | 100 | calcite | 0 |
| 18 | 360 | 0.36 | 100 | calcite | 0 |
| 19 | 740 | 0.19 | 50 | calcite | 5 |
| 20 | 380 | 0.10 | 50 | calcite | 5 |
| 21 | 300 | 1.20 | 200 | NH$_4$Cl | 52 |
| 22 | 340 | 1.36 | 200 | NH$_4$Cl | 65 |
| 23 | 340 | 1.36 | 200 | NH$_4$Cl | 29 |
| 24 | 320 | 1.28 | 200 | NH$_4$Cl | 48 |
| 26 | 380 | 1.52 | 200 | NH$_4$Cl | 108 |
| 27 | 300 | 1.20 | 200 | NH$_4$Cl | 46 |
| 28 | 400 | 1.60 | 200 | dry | 0 |
| 29 | 300 | 0.68 | 150 | dry | 0 |
| 30 | 300 | 0.68 | 150 | NH$_4$Cl | 15 |
| 31 | 300 | 0.68 | 150 | NH$_4$Cl | 40 |
| 32 | 300 | 0.68 | 150 | NH$_4$Cl | 12 |
| 33 | 360 | 0.81 | 150 | NH$_4$Cl | 23 |
| 34 | 340 | 0.77 | 150 | NH$_4$Cl | 5 |
| 35 | 390 | 0.88 | 150 | NH$_4$Cl | 42 |
| 37 | 360 | 0.36 | 100 | NH$_4$Cl | 25 |
| 38 | 340 | 0.34 | 100 | NH$_4$Cl | 16 |
| 40 | 300 | 0.08 | 50 | NH$_4$Cl | 5 |

Table 2. Experimental conditions for high resolution PSC experiments with continuous deformation recording

| Nb. | Fluid chemistry | Dead-weight (g) |
|---|---|---|
| 1 | saturated aqueous solution of calcite | 0; 80; 167 |
| 2 | saturated aqueous solution of calcite | 167 |
| 3 | dry (dried with nitrogen, immersed in hexadecane) | 167 |

Table 3. Rates of deformation at different time intervals of the first high resolution PSC experiment (see Table 2)

| time (h) | indentation rate (nm/h) | dead weight (g) |
|---|---|---|
| 9-15 | 23.3 | 80 |
| 19-24 | 46.6 | 80 |
| 30-35 | 15.0 | 80 |
| 48-54 | 27.6 | 167 |
| 69-79 | 18.3 | 167 |
| 79-89 | 14.6 | 167 |
| 109.5-113.3 | 0.0 | 0 |
| 122-128 | 15.4 | 167 |
| 169-182 | 6.5 | 167 |